# Magnetic and transport properties of electron-doped superconducting thin films: Pairing symmetry, pinning and spin fluctuations

S. Sergeenkov[1], A.J.C. Lanfredi[2] and F.M. Araujo-Moreira[2]

[1]Departamento de Física, CCEN, Universidade Federal da Paraíba, Cidade Universitária, 58051-970 João Pessoa, PB, Brazil

[2]Grupo de Materiais e Dispositivos, Centro Multidisciplinar para o Desenvolvimento de Materiais Cerâmicos, Departamento de Física, Universidade Federal de São Carlos, 13565-905, São Carlos, SP, Brazil

## Abstract

In this Chapter we review our latest results on magnetic (AC susceptibility) and transport (resistivity) properties of $Pr_{1.85}Ce_{0.15}CuO_4$ (PCCO) and $Sm_{1.85}Ce_{0.15}CuO_4$ (SCCO) thin films grown by pulsed laser deposition technique. Three main topics of our studies will be covered.
We start with a thorough discussion of the pairing symmetry mechanisms in optimally-doped SCCO thin films based on the extracted with high accuracy temperature profiles of penetration depth $\lambda(T)$ using a high-sensitivity home-made mutual-inductance technique. In particular, we found that above and below a crossover temperature $T^*=0.22\ T_C$, our films are best-fitted by a linear and quadratic dependencies, respectively, with physically reasonable values of d-wave node gap parameter $\Delta$ and paramagnetic impurity scattering rate $\Gamma$.
Our next topic is related to the flux distribution in our films. More precisely, we present a comparative study on their pinning ability at low magnetic fields extracted from their AC susceptibilities. Depending on the level of homogeneity of our films, two different types of the irreversibility line (IL), $T_{irr}\equiv T_p(H)$, defined as the intergrain-loss peak temperature in the imaginary part of susceptibility and obeying the law $1-T_p/T_C \propto H^q$, have been found. Namely, more homogeneous PCCO films (with grain size of the order of 2μm) are found to be best-fitted with q=2/3 while less homogeneous SCCO films (with grain size of the order of 500 nm) follow the IL law with q=1. The obtained results are described via the critical-state model taking into account the low-field grain-boundary pinning.
And finally, to emphasize non-trivial transport properties of electron-doped superconductors, we demonstrate our recent results on the temperature behavior of resistivity $\rho(T)$ for the high-quality optimally-doped SCCO thin films, paying special attention to their normal state properties. In addition to the expected contributions from the electron-phonon and electron-electron scattering processes, we also observed an unusual step-like behavior of $\rho(T)$ around T=87K very similar to the one seen in inelastic neutron scattering data. Given that Sm has a larger ion size than Pr and assuming that the long-range AFM correlations should be even stronger in thin films (than in single crystals), we attribute the appearance of this kink in our SCCO films to the manifestation of thermal excitations due to spin fluctuations induced by $Sm^{3+}$ moments through $Cu^{2+}$-$Sm^{3+}$ interaction.

## I. Introduction

The accurate experimental determination of the temperature behavior of the magnetic penetration depth, λ(T), has been of great interest to the scientific community since the very discovery of high-$T_C$ superconductors. Since the effective value of λ(T) is extremely sensitive to local inhomogeneities, a thorough analysis of its low-temperature profile gives probably one of the most reliable methods to determine the quality of a superconducting material (especially in the form of thin films), which is of utter importance for applications [1,2].

On the other hand, the magnetic penetration depth is strongly sensitive to the variations of the macroscopic superconducting order parameter and therefore its study can give important information about both the symmetry of the superconducting state and the pairing mechanisms. It is well established that most of the conventional low-$T_C$ superconductors have *s*-wave pairing symmetry. As for high-$T_C$ cuprates, the study of pairing symmetry in these materials has been (and still remains) one of the most polemical and active fields of research over the last few years [2] and the experimental determination of the temperature dependence of λ has been one of the most common methods in these studies. In particular, a power-like dependence $T^n$ of the penetration depth at low temperatures clearly points at a nodal structure of the superconducting gap (as expected for strongly correlated materials) where the exponent n depends on the type of the node in the **k**-space. Most phase-sensitive measurements [3,4] have revealed that hole-doped high-$T_C$ cuprates with nearly optimal doping have predominantly $d_{x^2-y^2}$ pairing symmetry. Regarding the case of the hole-doped cuprate $YBa_2Cu_3O_{7-\delta}$, some groups have reported experimental evidences for a pairing symmetry transition from pure $d_{x^2-y^2}$ (for under-doped compositions) to a mixed-type $d+id_{xy}$ (for over-doped compositions) [5]. At the same time, for electron-doped cuprates, which have composition of the form $Ln_{2-x}Ce_xCuO_4$ (where Ln corresponds to Pr, Nd, or Sm), the pairing mechanisms are not yet fully understood [6-10]. For example, using the point contact spectroscopy technique, Biswas *et al.* [7] have found strong evidences in favor of *d*-wave pairing symmetry in under-doped (x ≈ 0.13) $Pr_{2-x}Ce_xCuO_4$ (PCCO). Further studies revealed [8] that the low temperature superfluid density of Ce-based magnetic superconductors varies quadratically with temperature in the whole range of doping, in agreement with the theoretical prediction for a *d*-wave superconductor with impurity scattering. In addition, remeasured [9] magnetic-field dependence of the low-temperature specific heat of optimally-doped (x=0.15) PCCO give further evidence in favor of *d*-wave-like pairing symmetry in this material at all temperatures below 4.5 K. And finally, the recent penetration depth measurements on $Sm_{1.85}Ce_{0.15}CuO_4$ (SCCO) single crystals [10] have indicated that this magnetic superconductor exhibits a rather strong enhancement of diamagnetic screening below 4 K most probably driven by the Neel transition of Sm sublattice due to spin-freezing of Cu spins.

## II. Magnetic penetration depth and pairing symmetry of electron-doped high-$T_C$ superconducting thin films

In this Section we study the influence of local inhomogeneities on low-temperature dependence of the magnetic penetration depth λ(T) in high-quality optimally-doped $Pr_{1.85}Ce_{0.15}CuO_4$ (PCCO) and $Sm_{1.85}Ce_{0.15}CuO_4$ (SCCO) thin films grown by the pulsed laser deposition (PLD) technique. The λ(T) profiles have been extracted from conductance-voltage data by using a highly-sensitive home-made mutual-inductance bridge.

The structural quality of our samples was verified through X-ray diffraction (XRD) and scanning electron microscopy (SEM) together with energy dispersive spectroscopy (EDS) technique. Both XRD spectra and SEM data reveal that PCCO films are of higher structural quality than SCCO films (see Figure 1).

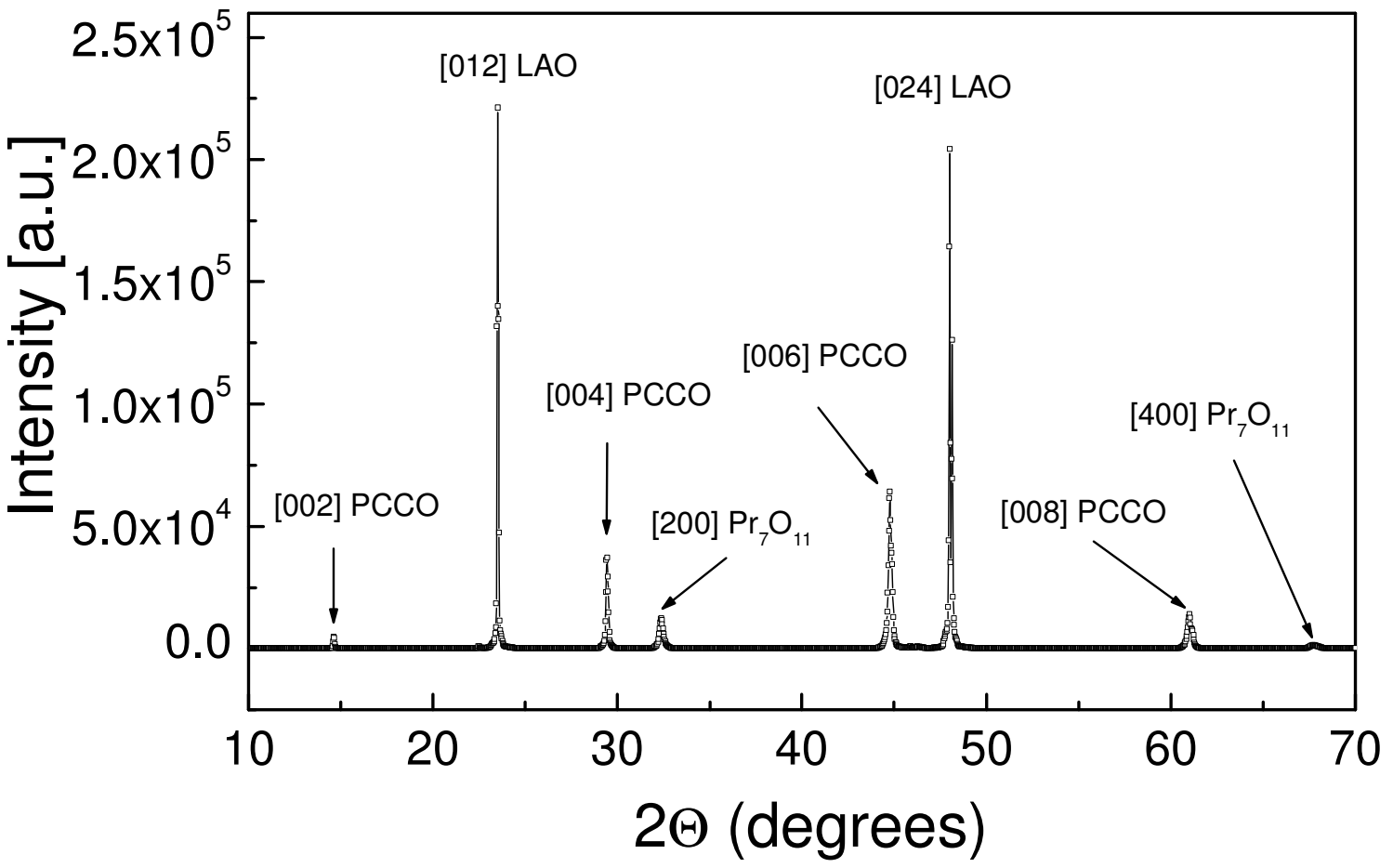

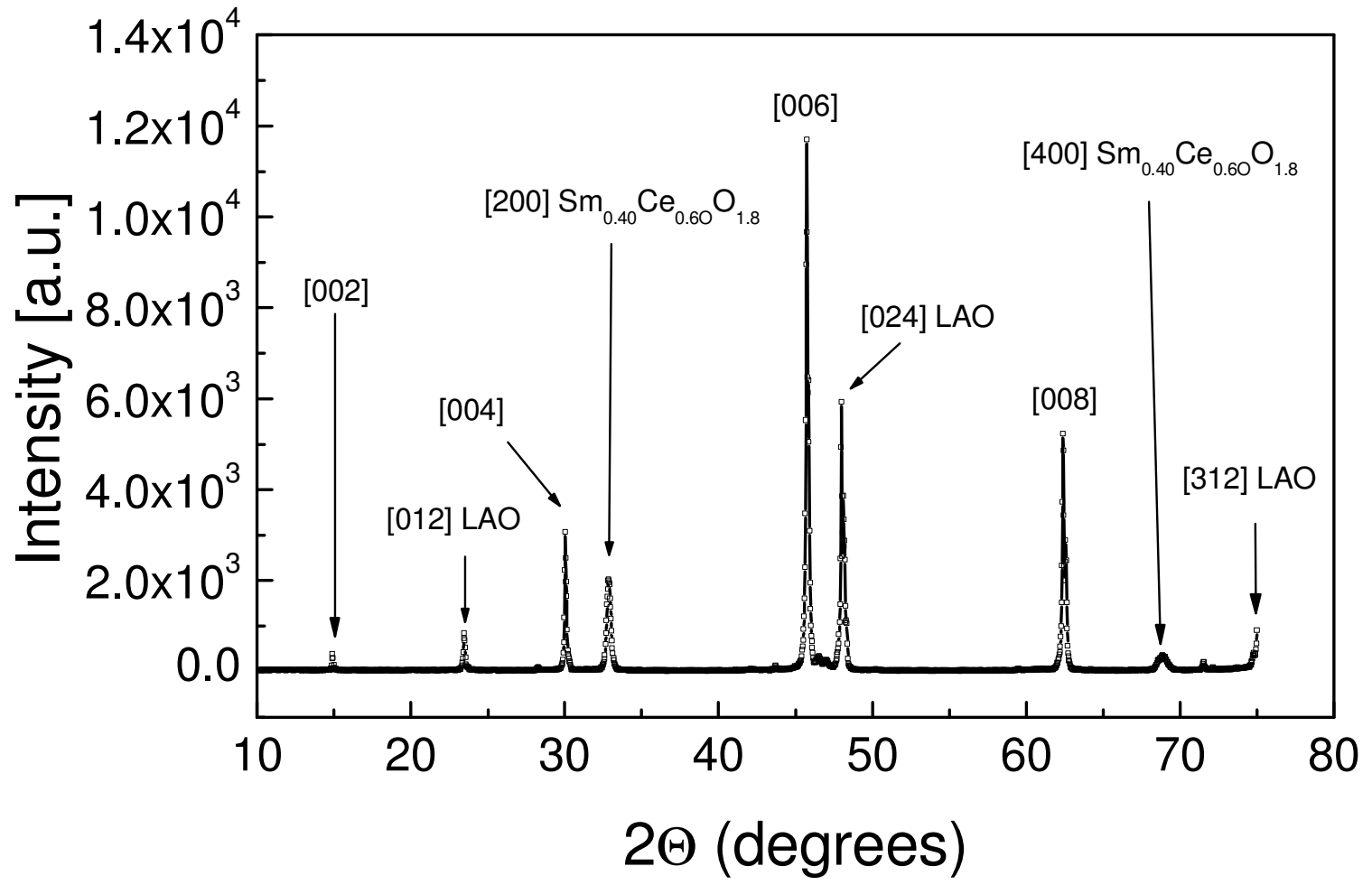

**Figure 1**. X-ray diffraction spectrum of PCCO (top) and SCCO (bottom) films.

The experimental bridge used in this work is based on the mutual-inductance method. To measure samples in the shape of thin films, the so-called *screening method* has been developed [11]. It involves the use of primary and secondary coils, with diameters smaller than the dimension of the sample. When these coils are located near the surface of the film, the response (i.e., the complex voltage output $V_{AC}$) does not depend on the radius of the film or its properties near the edges. In the reflection technique [12], an excitation (primary) coil coaxially surrounds a pair of counter-wound (secondary) pick-up coils. If we take the current in the primary coil as a reference, $V_{AC}$ can be expressed via two orthogonal components, i.e., $V_{AC} = V_L + iV_R$. The first one is the inductive component, $V_L$ (which is in phase with the time-derivative of the reference current) and the second one is the quadrature resistive component, $V_R$ (which is in phase with the reference current). It can be easily demonstrated that $V_L$ and $V_R$ are directly related to the average magnetic moment and the energy losses of the sample, respectively [13]. When there is no sample in the system, the net output from the secondary coils is close to zero because the pick-up coils are identical in shape but are wound in opposite directions. The sample is positioned as close as possible to the set of coils, to maximize the induced signal in the pick-up coils.

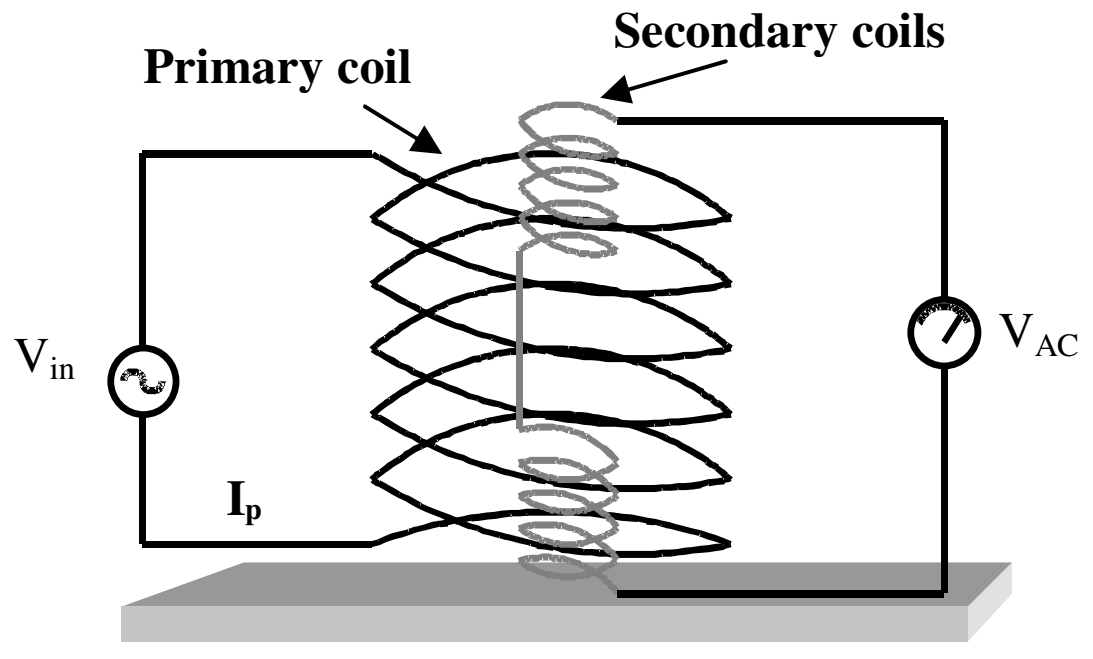

**Figure 2.** Sketch of the experimental bridge based on the mutual-inductance screening method.

An alternate current sufficient to create a magnetic field of amplitude $h_{AC}$ and frequency f is applied to the primary coil by an alternating voltage source, $V_{in}$. The output voltage of the secondary coils $V_{AC}$ is measured through the usual lock-in technique [14]. Figure 2 shows the sketch of the experimental bridge used in our study based on the mutual-inductance screening method.

To extract the profile of the penetration depth within the discussed here method, one should resolve the following equation relating the measured output voltage $V_{AC}$ to the $\lambda(T)$ sensitive sample features [12]:

$$V_{AC} = V' + iV'' = i\omega I_P \int_0^\infty \frac{M(x)}{1 + \frac{2}{\mu_0 (h_P + h_S)} \cdot \frac{1}{i\varpi G} x} \cdot dx \qquad (1)$$

where $I_P$ and $\omega=2\pi f$ are respectively the amplitude and the frequency of the current in the primary coil, $h_P$ ($h_S$) is the distance from the primary (secondary) coil to the sample, G is the total conductance of the sample, and M(x) is a geometrical factor [12]. Since the total impedance of the sample is given by [15] $Z = R + i\omega L_K$ the expression for the sample's total conductance reads:

$$G = \frac{1}{R + i\omega L_K} \qquad (2)$$

Here $L_k$ and R are the kinetic inductance and the resistance of the sample, respectively. From the above equations it follows that by measuring $V_{AC}(T)$ we can numerically reproduce the temperature dependencies of both $L_k$ and R.

From the two-fluid model, the relation between $L_k$ and $\lambda(T)$ for thin films (with thickness $d \ll \lambda$) is given by [1,2,15]:

$$L_K = \mu_0 \lambda \coth\left(\frac{d}{\lambda}\right) \approx \mu_0 \lambda \left(\frac{\lambda}{d}\right) \qquad (3)$$

This expression will be used hereafter to obtain $\lambda(T)$ from the measured $L_K(T)$ dependence. Instead of the tabulation based procedure used before [12], in the present study we have simultaneously determined G(T) from Eq.(1) and extracted both R(T) and $L_K(T)$ using Eq.(2). Then from the temperature dependence of $L_K$ we recover the temperature dependence of $\lambda$. Fig. 3 presents the temperature behavior of the typical output voltages of the secondary coils, $V_{AC}$, measured for superconducting thin films under an alternate magnetic field of amplitude $h_{AC}$=100 mOe and frequency f=55 kHz for our SCCO sample. Typical results for extracted variation of $\lambda^2(T)/\lambda^2(0)$ for both SCCO and PCCO thin films are shown in Fig. 4.

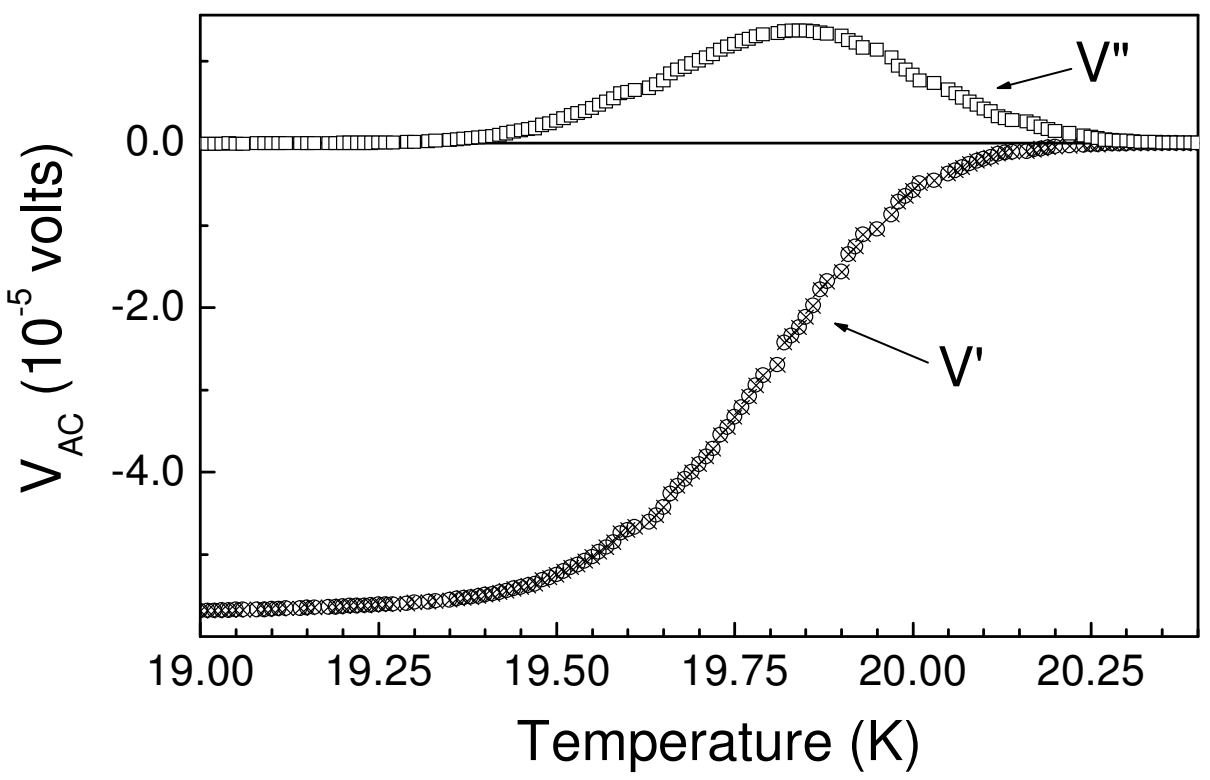

**Figure 3.** Temperature behavior of the typical output voltages of the secondary coils, $V_{AC}$, measured for superconducting thin films under an alternate magnetic field of amplitude $h_{AC}$=100 mOe and frequency f=55 kHz for SCCO sample ($T_C$ = 20.2 K) .

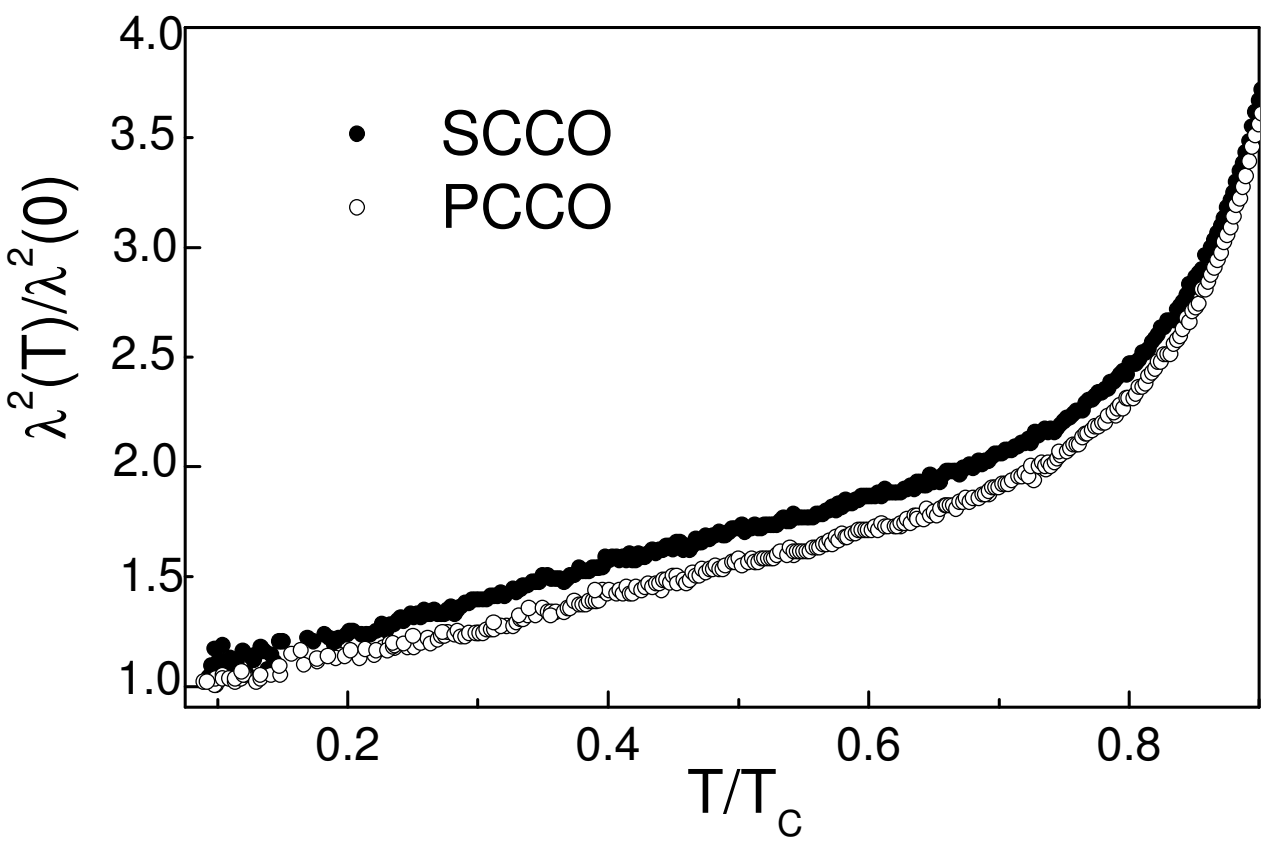

**Figure 4.** Extracted variation of $\lambda^2(T)/\lambda^2(0)$ as a function of the reduced temperature, obtained from Eqs.(1)-(3) for PCCO ($T_C$ =22.4 K) and SCCO ($T_C$ =20.2 K) thin films.

Turning to the discussion of the obtained results, recall [1] that for conventional BCS-type superconductors with *s*-wave pairing symmetry the superfluid fraction $x_S(T) = \lambda^2(0)/\lambda^2(T)$ saturates exponentially as T approaches zero. On the other hand, for a superconductor with a line of nodes, $x_S(T)$ will show a power-like behavior at low temperatures. In particular, the simple $d_{x^2-y^2}$ pairing state gives a linear dependence [16] $\Delta\lambda(T)/\lambda(0) \propto T$ for the low-temperature variation of in-plane penetration depth $\Delta\lambda(T)=\lambda(T)-\lambda(0)$. At the same time, in the presence of strong enough impurity scattering the linear T dependence changes to a quadratic $T^2$ dependence [17-20] $\Delta\lambda(T)/\lambda(0)\propto T^2$.

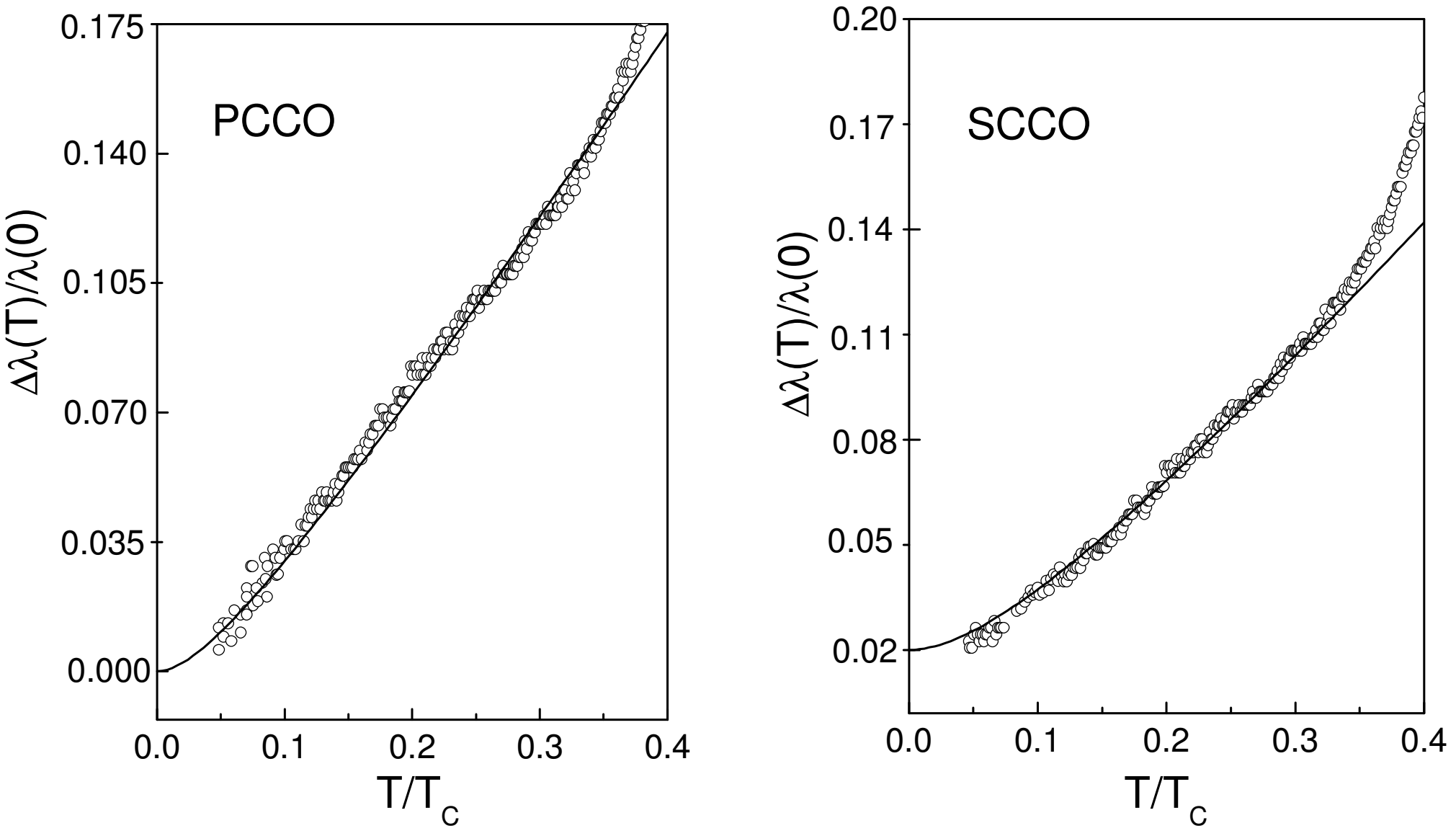

**Figure 5.** Low temperature fits (solid lines) of the extracted variation of the penetration depth $\Delta\lambda(T)/\lambda(0)$ in PCCO (left) and SCCO (right) thin films using the Goldenfeld-Hirschfeld interpolation formulae.

By trying many different temperature dependencies (including both exponential and power-like), we found that both our samples are best-fitted (see Fig. 5) by the so-called Goldenfeld-Hirschfeld interpolation formulae [20] $\Delta\lambda(T)/\lambda(0)=AT^2/(T+T_0)$ which describes a crossover between linear and quadratic dependencies above and below some temperature $T_0$. Here $A=\ln(2)k_B/\Delta_0$ with $\Delta_0$ being the amplitude of the zero-temperature value of the *d*-wave gap parameter, and the crossover temperature $T_0$ depends on the (unitary limit) scattering rate $\Gamma$ (which is proportional to the impurity concentration of the sample) as follows $T_0=\ln(2)k_B\Gamma^{1/2}\Delta_0^{\ 1/2}$. The fitting parameters are given in Table 1. Noticeably, the crossover temperature $T_0$ is lower for high-quality PCCO films ($T_0/T_C$ =0.13). In turn, this observation correlates well with a lower value of impurity scattering rate (in dimensionless units, $k_B^3\Gamma/T_C = 0.017$). Notice that the above estimates are in good agreement with the known results for high-quality PCCO thin films [8]. On the other hand, a less homogeneous SCCO film (with $T_C$=20.2K) exhibits a much stronger impurity scattering with the rate $k_B^3\Gamma/T_C = 0.062$ (starting to dominate below $T_0/T_C$ =0.26).

**Table 1.** Fitting parameters for temperature dependencies of penetration depth variations $\Delta\lambda(T)/\lambda(0)$ extracted from PCCO and SCCO thin films (see Fig. 5) along with the estimates for the nodal gap parameter $\mathbf{\Delta_o}$ and impurity scattering rate $\Gamma$ (in dimensionless units).

| **film** | **$T_C$ (K)** | **A $T_C$** | **$T_o/T_C$** | **$\Delta_o/k_BT_C$** | **$\Gamma k_B^3/T_C$** |
|---|---|---|---|---|---|
| PCCO | 22.4 | 0.35 | 0.13 | 2.0 | 0.017 |
| SCCO | 20.2 | 0.33 | 0.26 | 2.1 | 0.062 |

# III. Irreversibility line and low-field grain-boundary pinning in electron-doped superconducting thin films

The measurement of AC magnetic susceptibility still remains one of the most powerful methods to obtain important information on dissipation mechanisms in high-$T_C$ superconductors (HTS). To get useful information from such experiments, however, very careful control of sample's microstructure is required. While in high enough magnetic fields the dissipation is known to be dominated by flux motion of Abrikosov vortices [21-24], the low-field dissipation mechanisms (especially, in inhomogeneous and granular superconductors) are less obvious due to the numerous grain-boundary related effects which are better treated by the Josephson physics [25-27].

In this Section we present a comparative study of low-field AC magnetic susceptibility data on more homogeneous $Pr_{1.85}Ce_{0.15}CuO_4$ (PCCO) and less homogeneous $Sm_{1.85}Ce_{0.15}CuO_4$ (SCCO) thin films. The main idea of the experiments here reported is to study the influence of inhomogeneity on the dissipative properties of electron-doped thin films via the behavior of the irreversibility line (IL), $T_{irr} \equiv T_p(H)$, defined as the intergrain-loss peak temperature in the imaginary part of susceptibility $\chi''$ (T,H). This influence was found to result in a much higher pinning ability of less homogeneous SCCO thin films obeying the IL law $1- T_p/T_C \propto H^q$ with q=1 as compared to more homogeneous PCCO films with flux-creep exponent q=2/3.

A few PCCO and SCCO thin films (d=200nm thick) grown by pulsed laser deposition on standard $LaAlO_3$ substrates were used in our measurements (for discussion on different preparation techniques and chemical phase diagrams of electron-doped superconducting materials, including polycrystalline samples, single crystals, and thin films, see, e.g., [28-32] and further references therein). All samples showed similar and reproducible results. The SEM experiments reveal that PCCO films are of higher structural quality (more homogeneous) than SCCO films which show a pronounced granular structure (see Figure 6).

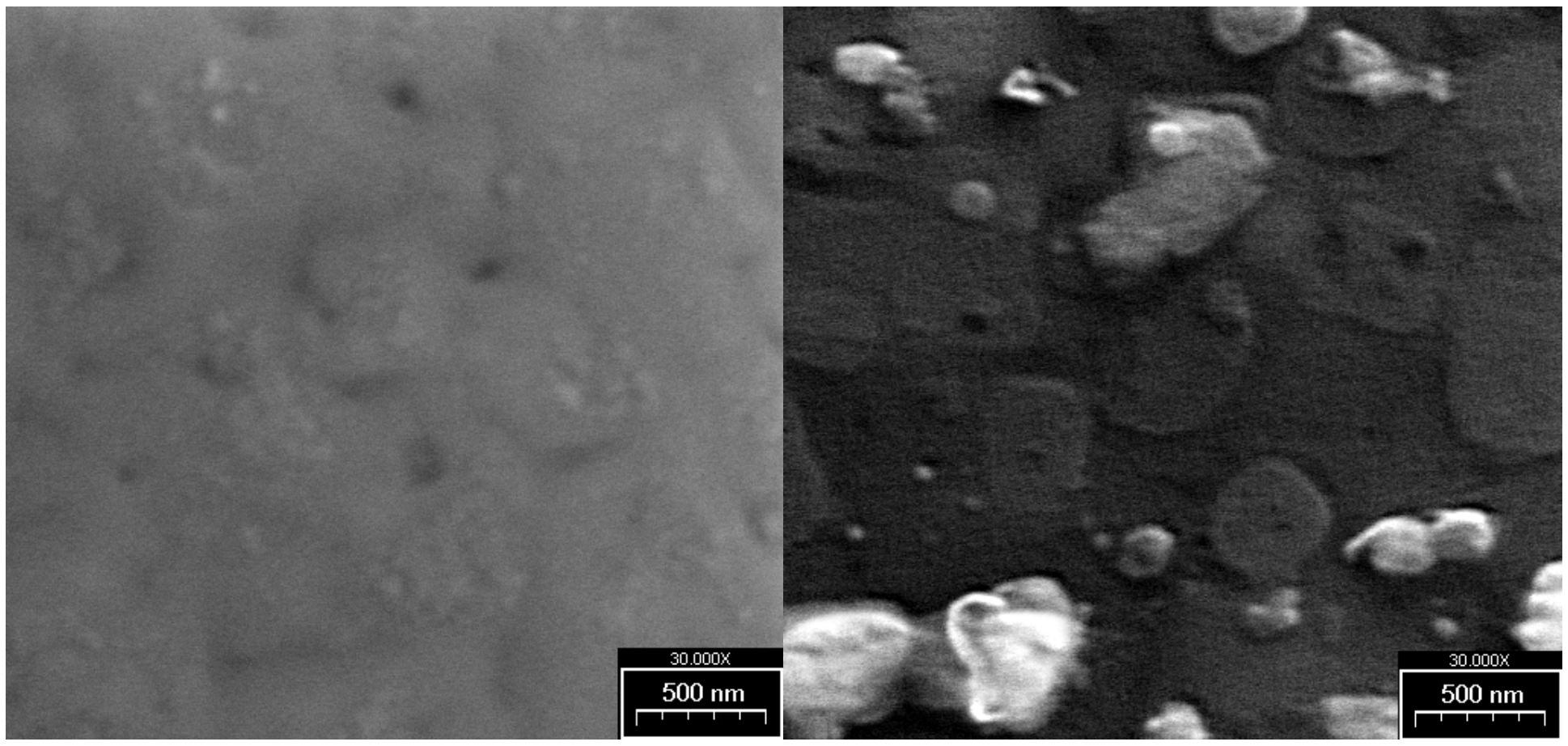

**Figure 6**. SEM scan photography of PCCO (left) and SCCO (right) samples (magnification 30000 times).

The average grain size in typical PCCO and SCCO films is estimated to be around 2μm and 0.5μm, respectively. Measurements of the real (χ') and imaginary (χ") parts of AC susceptibility were performed by using a MPMS magnetometer from the Quantum Design equipped with AC modulus [13,14,33,34]. All data are chosen from samples with the same dimensions and well placed parallel to the field in order to decrease the demagnetization correction. The symbol size used for data presentation takes into account error bars based on the temperature stability, reproducibility, and system precision. To account for a possible magnetic response from substrate, we measured several stand alone pieces of the substrate. No tangible contribution due to magnetic impurities was found. A typical temperature behavior of the measured complex AC magnetic susceptibility in PCCO and SCCO films in a small magnetic field (of amplitude $h_{AC}$=1.0Oe and frequency $f_{AC}$ =1.0kHz) is shown in Fig.7. The field dependence of the imaginary part χ" of the AC susceptibility in both films for the temperatures close to $T_C$ is depicted in Fig.8.

Due to small values of the applied magnetic field, it is natural to associate the peak temperatures $T_p(H)$ in Fig.8 with *intergrain* losses. The extracted values of the irreversibility temperature $T_p(H)$ for both samples are shown in Fig.9 in the form of the log-log plots. As is seen, more homogeneous PCCO films are well-fitted by the flux-creep mediated IL obeying the law $1 - T_p/T_C \propto H^q$ with q=2/3 while less homogeneous SCCO films (with grain size of the order of 500 nm) follow the IL law with q=1.

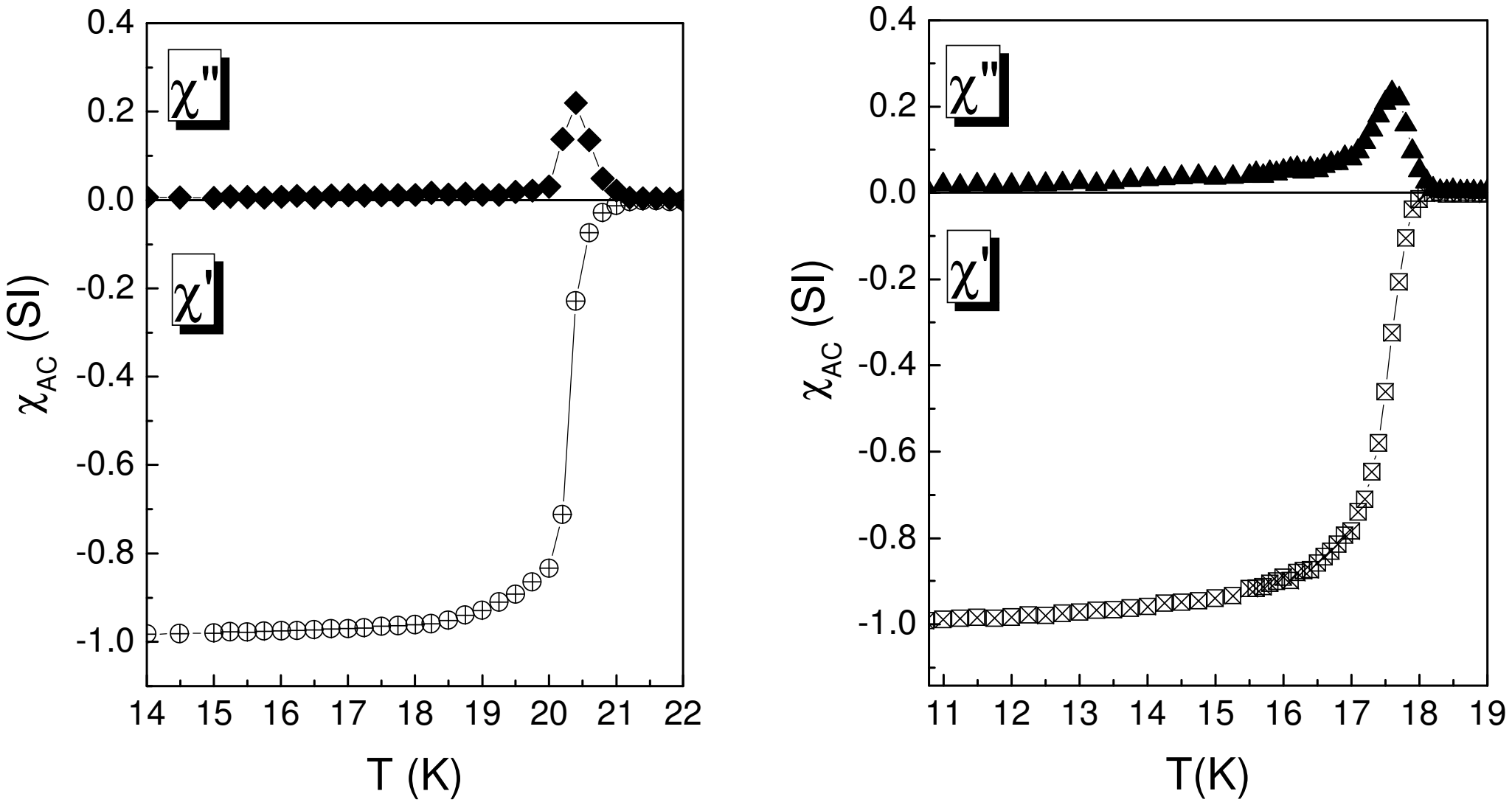

**Figure 7.** Temperature behavior of the AC susceptibility measured on PCCO (left) and SCCO (right) thin films for magnetic field of amplitude $h_{AC}$=1.0 Oe and frequency $f_{AC}$ =1.0 kHz.

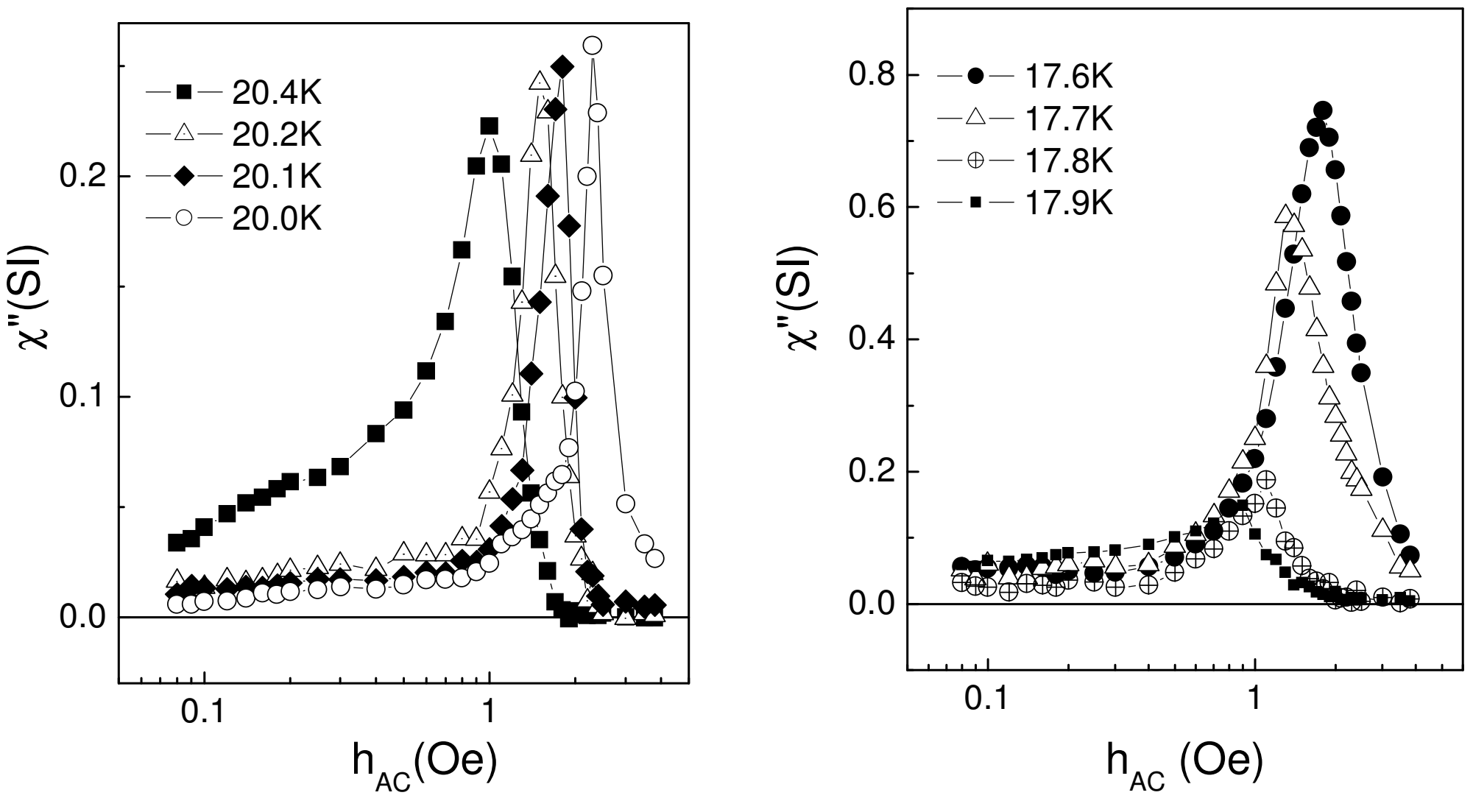

**Figure 8.** Magnetic field behavior of the imaginary part of AC susceptibility measured on PCCO (left) and SCCO (right) superconducting thin films at different temperatures.

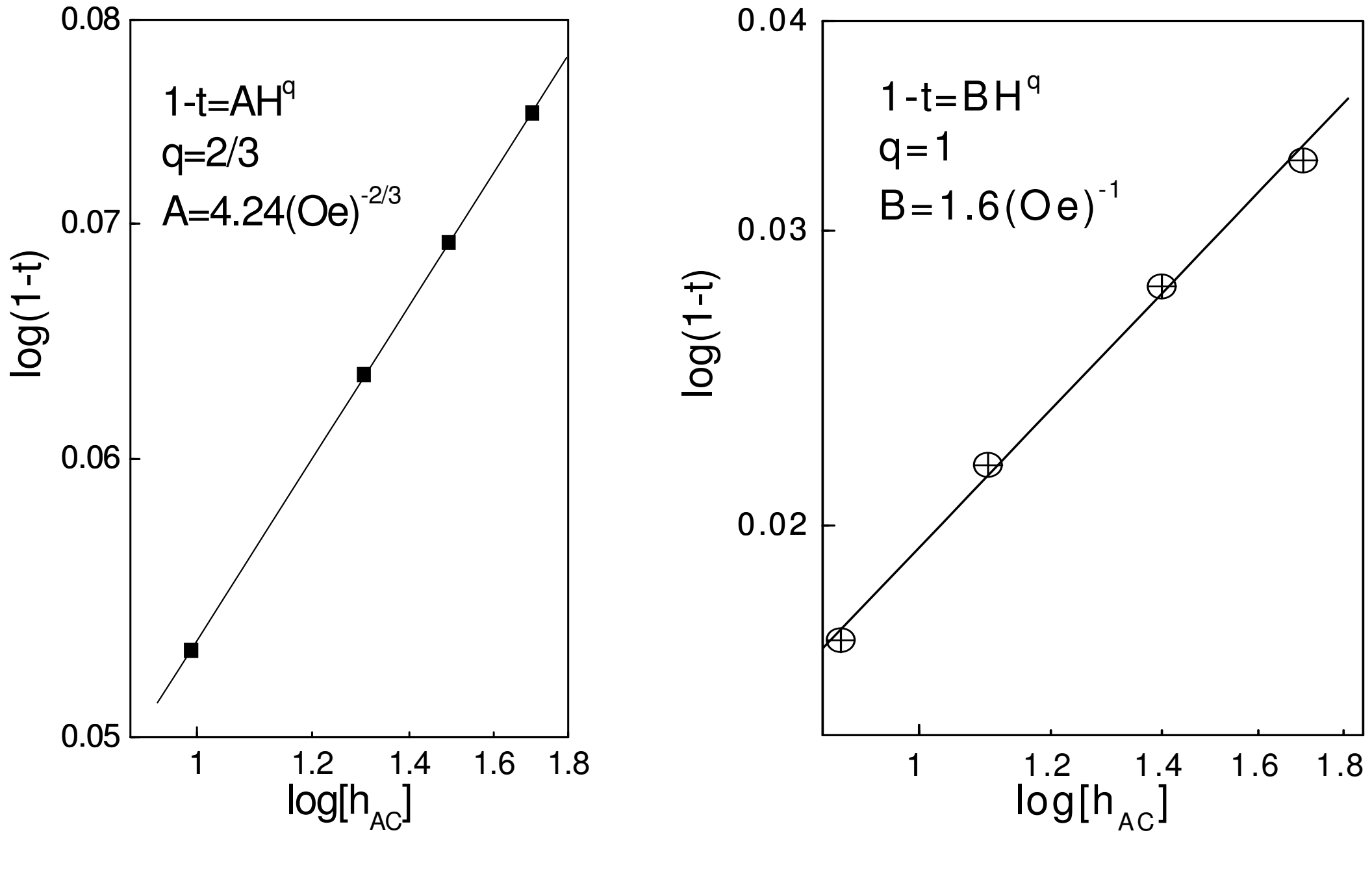

**Figure 9.** Log-log plot of the irreversibility lines $1\text{-}t \propto H^q$ (extracted from AC susceptibility data shown in Figure 8) for PCCO (left) and SCCO (right) films. Solid lines are the best fits according to Eqs.(4)-(8).

To interpret the above findings, we follow Müller's approach [35] (based on the Kim-Anderson critical-state model [36]) according to which the low-field dependence of the IL temperature $T_p(H)$ is governed by the following implicit equation (hereafter $H \equiv h_{AC}$)

$$\left[1+\frac{H}{H_C(T_p)}\right]^2 = 1+\frac{2d\mu_J(T_p)}{\mu_0\mu_{eff}(T_p)\left[H_C(T_p)\right]^2} \qquad (4)$$

where

$$\mu_{eff}(T) = \frac{2I_1(R/\lambda)\lambda}{I_0(R/\lambda)R} \qquad (5)$$

Here, $\lambda(T)$ is the London penetration depth, R is the average grain size, d is the film thickness, $\mu_{eff}(T)$ is the effective permeability of granular film, $H_C(T)$ is the characteristic field (see below), $\mu_J(T)$ is the so-called pinning-force density, and $I_0$ and $I_1$ are modified Bessel functions of the first kind. Notice that Eq.(4) is valid for applied fields larger than the lower Josephson field $H_C(T) = \frac{\phi_0}{4\pi\mu_0\lambda(T)R}$ when vortices nucleate along grain boundaries. These intergranular Josephson vortices are imbedded into a diamagnetic medium with effective permeability $\mu_{eff}(T)$ whose temperature dependence, in view of Eq.(5), is governed by the London penetration depth $\lambda(T) = \frac{\lambda(0)}{\sqrt{1-T/T_C}}$. The observed difference in behavior of IL is attributed to difference in average grain sizes in PCCO and SCCO films which, according to SEM scans (see Figure 6) are around R=2μm and R=500 nm, respectively.

Taking into account the explicit temperature dependence of the pinning-force density within the grain-boundary pinning model [37]

$$\mu_J(T) = \mu_J(0)(1 - T/T_C)^{3/2} \qquad (6)$$

we propose the following scenario for the observed IL behavior.

Since near $T_p$ in more homogeneous PCCO films (see Fig.6a) $R > \lambda(T)$, and hence $\mu_{eff}(T) \approx 2\lambda(T)/R$, from Eq.(4) we find the usual flux-creep dominated law (see Fig.9a)

$$1 - \frac{T_p}{T_C} = AH^{2/3} \quad \text{with} \quad A = \left[\frac{2\mu_0\lambda(0)H_C(0)}{\mu_J(0)dR}\right]^{2/3} \qquad (7)$$

On the other hand, in more granular SCCO films (see Fig.6b) near $T_p$ we have the opposite situation since in this case $R < \lambda(T)$, and hence $\mu_{eff}(T) \approx 1$. As a result, Eqs.(4)-(6) bring about the observed linear behavior of the IL (see Fig.9b)

$$1 - \frac{T_p}{T_C} = BH \quad \text{with} \quad B = \frac{\mu_0 H_C(0)}{\mu_J(0)d} \qquad (8)$$

By calculating the coefficients A and B from the IL curve slopes on a log-log plot, we can estimate the pinning-force densities $\mu_J(0)$ for both materials. Using for the film thickness d=200nm, London penetration depths [10] $\lambda_P(0)$=250nm, $\lambda_S(0)$=500nm, and average grain sizes R=2μm and R=0.5μm, from Eqs.(7) and (8) we obtain $\mu_{JP}(0)=3\text{x}10^4$TA/m$^2$ and $\mu_{JS}(0)=1.2\text{x}10^5$TA/m$^2$ for the pinning-force densities of PCCO and SCCO films, respectively. As expected, the above pinning values are larger than those seen in bulk granular materials [25-27]. Thus, for small applied magnetic fields, the flux pinning is dominated by the so-called electromagnetic pinning scenario characterized by the London pentration depth rather than coherence length (the latter is responsible for the so-called core pinning scenario in high enough magnetic fields). Within this scenario, the observed higher pinning ability of SCCO films near $T_p$ can be attributed to a perfect match between the average grain size R and the correspondent London penetration depth $\lambda_S(T_p)$. While in the case of a more homogeneous PCCO film (with the average grain size of R=2000nm) the ratio $\lambda_P(T_p)$/R is much less optimal leading to a lower pinning ability of these films. And finally, it is instructive to point out that the obtained here results on low-field irreversibility lines in our granular films (governed by grain-boundary pinning of coreless Josephson vortices) principally differ from the high-field irreversibility lines observed in electron-doped single crystals (dominated by core pinning of Abrikosov vortices, including particular scenarios for melting of the vortex lattice) [24].

## IV. Possible manifestation of spin fluctuations in the temperature behavior of resistivity in $Sm_{1.85}Ce_{0.15}CuO_4$ thin films

Despite numerous investigations on many different physical properties of electron-doped superconductors (EDS), these interesting materials continue to attract attention of both experimentalists and theoreticians alike, especially as far as their low-temperature anomalies are concerned (see, e.g.,[38-42] and further references therein). Of particular interest is Sm-based EDS. Since Sm has a larger ion size than Ce, Pr and Nd, it is expected that paramagnetic scattering contribution to low-temperature behavior of $Sm_{2-x}Ce_xCuO_4$ should be much stronger than in $Pr_{2-x}Ce_xCuO_4$ and $Nd_{2-x}Ce_xCuO_4$. It should be mentioned also that in addition to their unusual pairing properties, EDS exhibit some anomalous normal state behavior far above $T_C$ with a noticeable presence of both electron-phonon and electron-electron contributions [43-45]. Recent inelastic neutron scattering experiments [46,47] on low-energy spin dynamics (for the energy spectrum ranging from 1meV to 10meV) in $LaPr_{0.88}Ce_{0.12}CuO_4$ (PLCCO) clearly demonstrated the evolution of PLCCO from nonsuperconducting antiferromagnet (with the Neel temperature $T_N = 210K$) to optimally doped superconductor (with $T_C = 24K$). Besides, a step-like intensity increase was observed at about $T_{sf} = 80K$ and linked to the manifestation of low-energy ($\hbar\omega_{sf} = k_B T_{sf} \approx 6.5meV$) long-range antiferromagnetic (AFM) spin fluctuations in the excitation spectrum induced by $\mathrm{Pr}^{3+}$ moments through $Cu^{2+}-\mathrm{Pr}^{3+}$ interaction [48].

In this Section we present our latest results on the temperature behavior of resistivity $\rho(T)$ for the optimally-doped $Sm_{1.85}Ce_{0.15}CuO_4$ films [42], paying special attention to their normal state properties. In addition to the expected contributions from the electron-phonon and electron-electron scattering processes, we also observed an unusual kink like behavior of $\rho(T)$ around T=87K very similar to the one seen in inelastic neutron scattering data [46,47]. Given that Sm has a larger ion size than Pr and assuming that the long-range AFM correlations should be even stronger in thin films (than in single crystals), we attribute the appearance of this kink in our SCCO films to the manifestation of thermal excitations due to spin fluctuations induced by $Sm^{3+}$ moments through $Cu^{2+}-Sm^{3+}$ interaction.

A few SCCO thin films (d=200nm thick) grown by pulsed laser deposition on standard $LaAlO_3$ substrates were used in our measurements (for more details on our samples including their other physical properties, see [42]). All samples showed similar and reproducible results. The structural quality of the samples was verified through X-ray diffraction (see Fig.1) and scanning electron microscopy together with energy dispersive spectroscopy technique. To account for a possible magnetic response from substrate, we measured several stand alone pieces of the substrate. No tangible contribution due to magnetic impurities was found. The electrical resistivity $\rho(T)$ was measured using the conventional four-probe method. To avoid Joule and Peltier effects, a dc current I=1mA was injected (as a one second pulse) successively on both sides of the sample. The voltage drop V across the sample was measured with high accuracy by a KT256 nanovoltmeter.

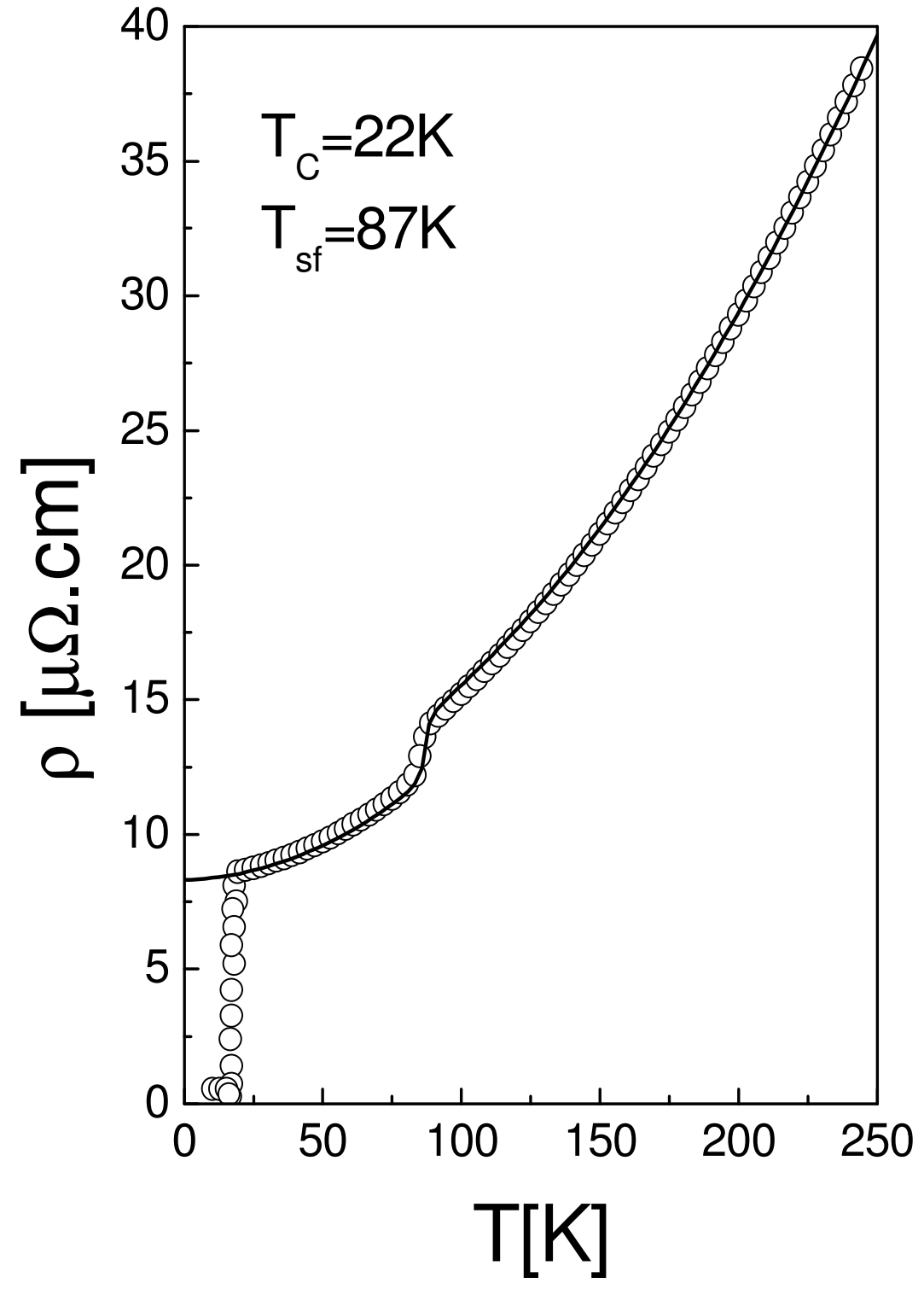

**Figure 10.** Temperature dependence of the resistivity $\rho(T)$ measured for a typical SCCO thin film. The solid line is the best fit according to Eq.(10).

Fig.10 shows the typical results for the temperature dependence of the resistivity $\rho(T)$ in our SCCO thin films. Quite a pronounced step (kink) is clearly seen around T=87K. Since, according to the X-ray diffraction spectrum (Fig.1), our films do not show any low-energy structural anomalies, it is quite reasonable to assume that the observed kink can be attributed to the manifestation of long-range AFM spin fluctuations induced by $Sm^{3+}$ moment with the characteristic energy $\hbar\omega_{sf} = 7meV$ corresponding to an effective temperature $T_{sf} = \hbar\omega_{sf} / k_B = 87K$ and a size of spin fluctuations domain $\xi_{sf} = \sqrt{\frac{\hbar}{2m\omega_{sf}}} \approx 2nm$.

More specifically, to account for fluctuation induced thermal broadening effects (of the width $\omega_{sf}$) we suggest a Drude-Lorentz type expression for this contribution (Cf. [49]):

$$\rho_{sf}(T) = \rho_{res} \int_{-\omega_{sf}}^{\Omega(T)-\omega_{sf}} \frac{\omega_{sf} d\omega}{\pi\left(\omega^2 + \omega_{sf}^2\right)} = \rho_{res}\left[\frac{1}{4} + \frac{1}{\pi}\tan^{-1}\left(\frac{T - T_{sf}}{T_{sf}}\right)\right] \quad (9)$$

where $\rho_{res} = \left(\omega_p^2 \varepsilon_0 \tau_0\right)^{-1}$ is the residual contribution with $\omega_p$ being the plasmon frequency, $\tau_0^{-1}$ the corresponding scattering rate, and $\varepsilon_0$ vacuum permittivity. Notice that $\rho_{sf}(0) = 0$.

The temperature dependence in Eq.(9) comes from the cutoff frequency $\Omega(T) = \frac{U(T)}{\hbar}$ which accounts for spin fluctuations with an average thermal energy $U(T) = \frac{1}{2} C < u^2 > \approx k_B T$ where [50] C is the force constant of a two-dimensional harmonic oscillator, and $< u^2 >$ is the mean square displacement of the magnetic Sm atoms from their equilibrium positions.

After trying many different temperature dependencies, we found that our SCCO films are rather well fitted (solid line in Fig.10) using the following expression for the observed resistivity:

$$\rho(T) = \rho_{res} + \rho_{sf}(T) + \rho_{e-ph}(T) + \rho_{e-e}(T) \quad (10)$$

where other two terms in the rhs of Eq.(10) are related, respectively, to electron-phonon contribution [43] $\rho_{e-ph}(T) = AT$ with $A = \frac{\lambda k_B}{\varepsilon_0 \hbar \omega_p^2}$ and to electron-electron contribution [44,45] $\rho_{e-e}(T) = BT^2$ with $B = \frac{k_B^2}{\varepsilon_0 \hbar \omega_p^2 E_F}$. Here, $\lambda$ is the electron-phonon coupling constant, and $E_F$ the Fermi energy. Using the experimentally found values of $\rho_{res} = 8.8\mu\Omega cm$, $A = 0.14\mu\Omega cm / K$, $B = 0.0012\mu\Omega cm / K^2$, and $T_{sf} = 87K$, the best fits through the data points produced $\omega_p = 2.1meV$, $\tau_0^{-1} = 9.5 \cdot 10^{-14} s^{-1}$, $\lambda = 1.2$, and $E_F = 0.2eV$ for very reasonable [43-45] estimates of the plasmon frequency, the impurity scattering rate, electron-phonon coupling constant, and the Fermi energy, respectively.

## V. Conclusion

In this Chapter we presented our latest results on magnetic and transport properties of electron-doped $Pr_{1.85}Ce_{0.15}CuO_4$ (PCCO) and $Sm_{1.85}Ce_{0.15}CuO_4$ (SCCO) thin films. Using a highly-sensitive home-made mutual-inductance technique associated with a new numerical procedure, we extracted with high accuracy the temperature profiles of penetration depths in optimally-doped PCCO and SCCO thin films. Based on the obtained results, we conclude that our findings confirm a universal pairing mechanism in electron-doped magnetic superconductors with *d*-wave nodal symmetry, and that deviations from the expected wave symmetry at the lowest temperatures are clear signals of structural inhomogeneity which can be tested via accurate measurement of the magnetic penetration depth. The values of the extracted impurity scattering rate $\Gamma$ were found to correlate with the quality of our samples. As expected, small (large) values of $\Gamma$ correspond to high (low) values of the critical temperature $T_C$ in more (less) homogeneous PCCO (SCCO) thin films. Furthermore, by analyzing the measured AC magnetic susceptibilities of PCCO and SCCO thin films as a function of temperature and magnetic-field strength, we associated the irreversibility line with the intergranular peaks in the imaginary part of AC susceptibilities. The obtained results are described in the framework of the Kim-Anderson critical-state model taking into account the grain-boundary pinning of Josephson vortices. And finally, we attributed an unusual kink like behavior observed in the temperature dependence of resistivity for our optimally-doped high-quality $Sm_{1.85}Ce_{0.15}CuO_4$ films around T=87K to a possible manifestation of thermal excitations due to spin fluctuations induced by $Sm^{3+}$ moments.

## Acknowledgements

We gratefully acknowledge financial support from Brazilian agencies CAPES and FAPESP.